\documentclass[twocolumn,superscriptaddress,showkeys,floatfix]{revtex4-1} 

\usepackage{amsfonts}
\usepackage{graphicx}
\usepackage{bm}
\usepackage{dcolumn}
\usepackage[colorlinks,linkcolor=blue,citecolor=blue]{hyperref}
\usepackage{epsfig}
\usepackage{multirow}
\usepackage{amsmath}
\usepackage{hhline}
\usepackage{mathptmx}
\usepackage{eucal}
\usepackage{color}
\usepackage{epstopdf}
\usepackage{natbib}
\usepackage{textcomp}

\usepackage[T1]{fontenc}
\usepackage[latin9]{inputenc}
\usepackage{amsmath,amssymb,amstext}
\usepackage{esint}
\usepackage{bm,bbold}
\DeclareMathOperator\Tr{Tr}
\DeclareMathOperator\Real{Re}

\begin{document}
\title{Revealing the magnetic proximity effect in EuS/Al bilayers through superconducting tunneling spectroscopy}

\author{E. Strambini}
\email{elia.strambini@sns.it}
\affiliation{NEST Istituto Nanoscienze-CNR  and Scuola Normale Superiore, I-56127 Pisa, Italy}
\author{V. N. Golovach}
\affiliation{Centro de Fisica de Materiales (CFM-MPC), Centro Mixto CSIC-UPV/EHU,Manuel de Lardizabal 5, E-20018 San Sebastian, Spain}
\author{G. De Simoni}
\affiliation{NEST Istituto Nanoscienze-CNR  and Scuola Normale Superiore, I-56127 Pisa, Italy}
\author{J. S. Moodera}
\affiliation{Department of Physics and Francis Bitter Magnet Lab, Massachusetts Institute of Technology, Cambridge, Massachusetts 02139, USA}
\author{F. S. Bergeret}
\email{sebastian_bergeret@ehu.eus}
\affiliation{Centro de Fisica de Materiales (CFM-MPC), Centro Mixto CSIC-UPV/EHU,Manuel de Lardizabal 5, E-20018 San Sebastian, Spain}
\affiliation{Donostia International Physics Center (DIPC),Manuel de Lardizabal 4, E-20018 San Sebastian, Spain}
\author{F. Giazotto}
\email{francesco.giazotto@sns.it}
\affiliation{NEST Istituto Nanoscienze-CNR  and Scuola Normale Superiore, I-56127 Pisa, Italy}

%
%
%


\begin{abstract}
\noindent
A ferromagnetic insulator in contact with a superconductor is known to induce exchange fields ranging from few to tens of Tesla driven splitting of the Bardeen-Cooper-Schrieffer (BCS) density of states singularity by a magnitude proportional to the magnetization, and the exchange field penetrating into the superconductor to a depth comparable with the superconducting coherence length. 
This long range magnetic proximity effect in EuS/Al bilayers and the exchange splitting of the BCS peaks position and intensity  were found to be influenced by the domain structure in EuS at the nanoscale present already in its unmagnetized state. Upon magnetizing the EuS the splitting was enhanced while peaks became symmetric. Conductance measurements as a function of bias voltage at the lowest temperatures could theoretically relate the line shape of the split BCS DoS with the characteristic domain structure in the ultra thin EuS layer. 
These results pave the way to engineering triplet superconducting correlations at domain walls in EuS/Al bilayers. 
Furthermore, the clear gap and splitting observed in our tunneling spectroscopy measurements show that it can be an excellent candidate for substituting strong magnetic fields in experiments studying Majorana bound states.
\end{abstract}

\maketitle
Europium sulfide is a classic Heisenberg ferromagnetic insulator (FI) with a Curie temperature of $16.7\,\textrm{K}$~\cite{mauger_magnetic_1986,Hao_Spin-filter_1990}, that exceeds the transition temperature of most of the conventional superconductors.  Together with EuO this material can be used as a very efficient spin-filter barrier \cite{moodera_phenomena_2007,moodera_frontiers_2010}. 
Experiments carried out in the eighties have demonstrated that the exchange field of FIs, such as EuS and EuO, can split the excitation spectrum of an adjacent superconductor (S), such as an Al thin film~\cite{Hao_Spin-filter_1990,Moodera_Electron-spin_1988,tedrow_spin-polarized_1986}.
This discovery opened up the way for performing spin-polarized tunneling measurements without the need of applying large magnetic fields~\cite{tedrow_spin-polarized_1986} -- a feature which is highly desirable when superconducting elements are present in the electronic circuit. More recently, EuS has  also been  used  to create  strong interfacial exchange fields in graphene \cite{wei_strong_2016}  and  topological insulators\cite{wei2013exchange}.


A renewed interest in studying Ferromagnetic/superconductor structures came with the development of superconducting spintronics~\cite{linder_superconducting_2015}.
The interaction between the superconducting condensate and the exchange field of a ferromagnet  creates triplet  superconducting pairs  which are able to carry non dissipative, spin-polarized currents \cite{bergeret_odd_2005}. 
The creation and control over the triplet correlations is intimately related with the magnetic configuration of the ferromagnet, with the domain walls playing an important role~\cite{bergeret_long-range_2001}.

In the case of ferromagnetic insulators, a series of interesting phenomena have been predicted to occur in S/FI structures with spin-split density of states (DoSs), such as huge thermoelectric effects~\cite{Ozaeta_Predicted_2014,Machon_Nonlocal_2013,Kolenda_Observation_2016,Giazotto_Proposal_2014,Giazotto_Very_2015} and highly efficient spin and heat valves~\cite{Huertas-Hernando_Absolute_2002,Giazotto_quantum_2013,Giazotto_Phase-tunable_2013,giazotto_huge_2006}.
These effects can be exploited for creation of spin-polarized currents with a high degree of polarization~\cite{Huertas-Hernando_Absolute_2002,Giazotto_Manipulating_2005,Giazotto_Superconductors_2008}, for on-chip cooling at the nanoscale~\cite{Giazotto_Opportunities_2006,kawabata_efficient_2013}, and for low-temperature thermometry and highly sensitive detectors and bolometers~\cite{giazotto_ferromagnetic-insulator-based_2015}.

A spin-split superconducting DoSs is also an essential ingredient in Majorana-based quantum computing~\cite{Aasen_milestones_2016,beenakker_road_2016}.
The exchange splitting of the BCS singularity observed in EuS/Al bilayers is as large as the splitting caused by an external magnetic field of several Tesla.
Therefore, replacing the superconductor by an
EuS/Al bilayer or another FI-S carefully-designed structure should allow one to reduce significantly or, in certain cases, even avoid the use of magnetic fields in experimental setups with Majorana fermions.
This possibility becomes especially attractive at the production cycle, since having to apply strong magnetic fields is impractical, whereas the magnetization of an island of FI can be manipulated on chip through an electric field or spin transfer torque.
%
%

\begin{figure*}[th!]
\centerline{\includegraphics[width=1.0\textwidth]{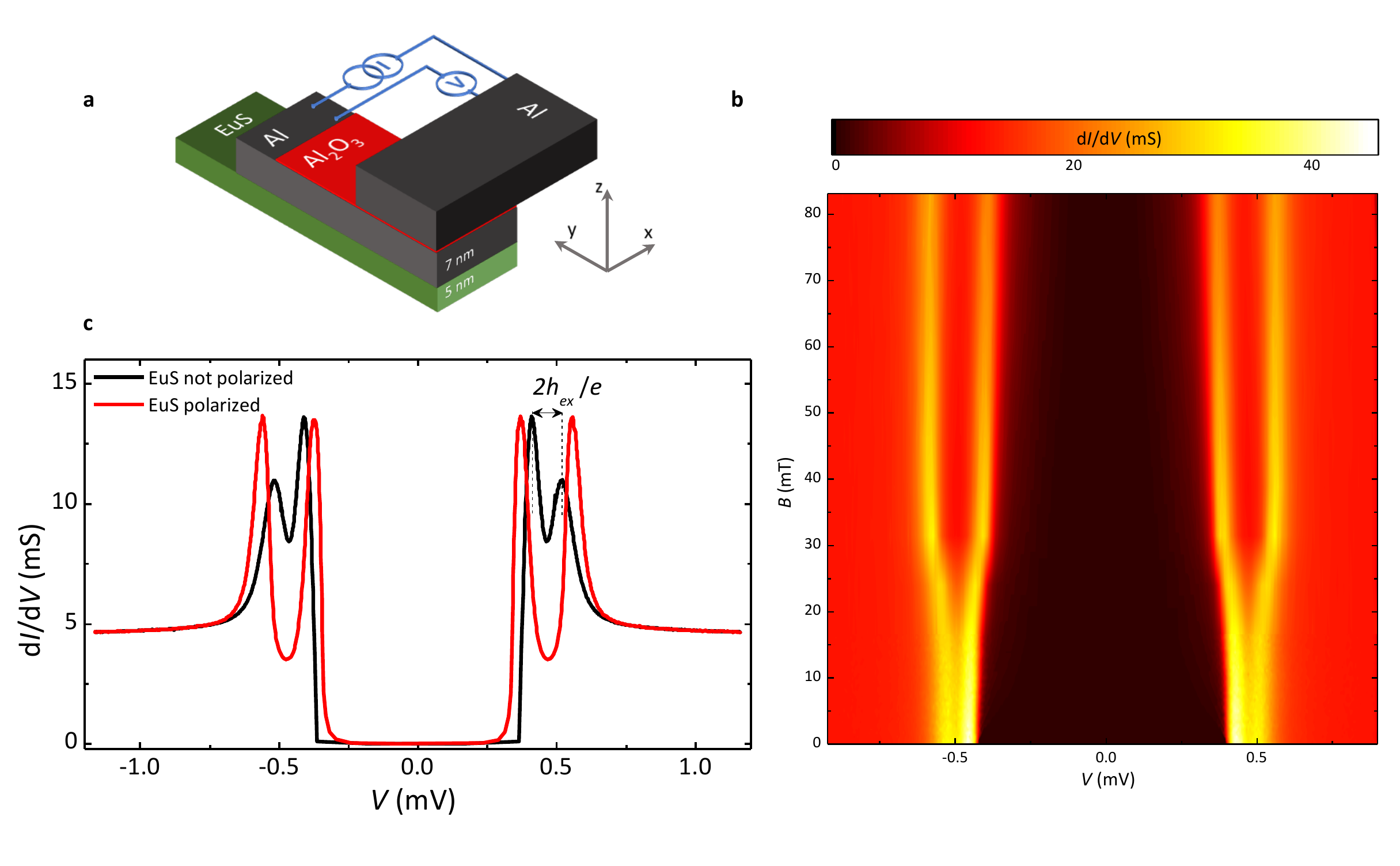}}
\caption{
\label{FirstMag}
\textbf{Junction layout and tunneling spectroscopy of the first magnetization.}
(\textbf{a}) Sketch of the cross bar forming the EuS(5)/Al(7)/Al$_{2}$O$_{3}$/Al(18) vertical tunnel junction (the thickness is in nanometers). The area of the junction is a square of 290x290 $\mu$m$^2$.
(\textbf{b}) Evolution of the differential conductance, obtained from the numerical derivative of the $I\textrm{-}V$ curves, as a function of the voltage drop ($V$) and in-plane magnetic field ($B$) during the first magnetization of the EuS layer.
(\textbf{c}) Comparison between the differential conductance of the tunnel junction measured at zero field before (black curve) and after (red curve) the magnetization of the EuS layer. All the measurements were taken at $25\,\textrm{mK}$.
}
\end{figure*}

All these applications need a sizable splitting of the superconducting DoSs in a large temperature range below the superconducting critical temperature.
A first and essential step towards controlling the magnitude of the exchange splitting is to understand the magnetic proximity effect induced by the FI material in an adjacent superconductor. Although different FI/S systems have been studied for almost three decades, there is still a great deal of controversy about the relation between the magnetic configuration of the EuS and the spin splitting induced in the superconductor~\cite{Hao_Spin-filter_1990,xiong_spin-resolved_2011}. Moreover, very few articles focus on the behavior of the EuS/Al bilayers at temperatures well below $1\,\textrm{K}$.


Here, we present an accurate  tunneling spectroscopy of the superconducting DoS of an EuS/Al bilayer in the temperature range $30\,\textrm{mK}$--$1.2\,\textrm{K}$.
The exchange splitting observed in the Al layer reaches up to $0.2\,\textrm{meV}$ in the presence of a moderate in-plane magnetic field of $30\,\textrm{mT}$, which is applied in order to align the magnetic domains of the EuS. Once magnetized, the spin-splitting is also clearly observed at zero applied field. 
Most notable, however, is the fact that the experimental data exhibits the  splitting even in the demagnetized phase of the EuS, \emph{i.e.}, even before the first application of a magnetic field. Moreover, the line shape of the BCS singularity is considerably reconstructed as compared to the standard BCS line shapes observed in the magnetically ordered state:  In a homogeneous exchange-split superconductor, the total DoS is a sum of a spin-up and a spin-down BCS DoS, shifted in energy with respect to each other by the exchange splitting, resulting in a four-peak structure with the outer peaks higher than the inner ones~\cite{tedrow_spin-polarized_1986}. However,  our measurements in the demagnetized phase of the EuS  shows that the peak heights  have the opposite asymmetry as compared to the homogeneous case.  

In order to understand the experimental observations, we model the EuS as a periodic structure of magnetic domains of different sizes and compute the DoS of the Al film with the help of the quasiclassical Green's function formalism. 
Our analysis shows that an exchange splitting in the DoS of the Al layer before the magnetization of the EuS can  only be obtained if the EuS layer consists predominantly of large domains, \emph{i.e.}\ much larger than the superconducting coherence length. Yet, the fact that the BCS singularity is considerably reshaped in the demagnetized case indicates that domain walls are not that rare, and contribute sizeably to the tunneling spectroscopy.  We identify the main physical processes responsible for the reconstruction of the BCS singularity around a domain wall, and make predictions about a possible scanning tunneling microscopy of the EuS/Al bilayer.

Further information about the magnetic configuration of EuS can be extracted from the temperature dependence of  the  exchange splitting. Surprisingly  there is a 10\% reduction of the  splitting  when the temperature is varied from 30 to 900 mK. We attribute this  large change of the splitting over a temperature range  much smaller  than the  Curie temperature  to the Al/EuS interface  that may consist of  single localized spins  (Eu  atoms) coupled to the EuS layer only by one bond. We support this hypothesis  by  a calculation of the average magnetic moment at the interface .

Finally we use the well-pronounced gap to achieve a large tunneling magnetoresistance (TMR) values at the magnetization reversal point $B_c\approx 18.5\,\textrm{mT}$, ranging from $200\,\%$ at $T=30\,\textrm{mK}$ up to $700\,\%$ at $T=850\,\textrm{mK}$. Notably this large TMR values are achieved using only one magnetic layer. Apart from serving as a measurement of the  figure of merit for the hardness of the gap in a functional FI/S device, the large observed TMR  suggest that Al/EuS systems can be used as building block for a superconducting spin based electronic devices. 


\section{Samples and measurements}
The tunneling spectroscopy of the EuS/Al bilayer has been done on EuS(5)/Al(7)/Al$_{2}$O$_{3}$/Al(18) tunnel junctions (thickness in nanometers). 
Samples consist of cross bars fabricated by electron-beam evaporation on in situ metallic shadow mask (see Methods for fabrication details). The typical area of the FI/S/I/S junction is 290x290 $\mu$m$^2$. The tunneling spectroscopy of the junctions is obtained by measuring the $V\textrm{-}I$ characteristics in a DC four-wire setup sketched in Fig.~\ref{FirstMag}a from which the differential conductance is evaluated via numerical differentiation. 
The cross bar junctions are characterized at cryogenic temperatures, down to $25\,\textrm{mK}$, in a filtered cryogen-free dilution refrigerator.

\section{Results}
Samples are first cooled down from room temperature to $30\,\textrm{mK}$ in a non-magnetic environment. 
Surprisingly, before the application of any magnetic field, the $dI/dV$ versus $V$ shows four clear peaks indicating an exchange splitting in the DoSs of the bottom Al layer (as shown in Fig.~\ref{FirstMag}b,c for two similar devices).
The symmetry and position of these peaks, in a first approximation, can be well described within the Tedrow and Meservey theory~\cite{tedrow_spin-polarized_1986} of quasiparticle spin-polarized tunneling, for which four superconducting sum gap peaks are expected at the voltages
\begin{equation}
e V_{\textrm{peak}} \simeq \pm  (\Delta_1+\Delta_2) \pm h_{\textrm{ex}},
\end{equation}
where $\Delta_i$ is the pairing potential of each superconductor forming the junction, $e$ is the electron charge, and $h_{\textrm{ex}}$ is the exchange energy induced in the bottom Al layer in contact with the EuS film.
Assuming equal pairing potentials in the two Al layers, the measurement is compatible with $\Delta\approx 230\,\mu\textrm{eV}$ and an exchange splitting $2 h_{\textrm{ex}} \approx 110\,\mu\textrm{eV}$. The latter is equivalent to an effective magnetic field $B_{\textrm{ex}} = 2 h_{\textrm{ex}} / \textsl{g}\mu_B  \sim 1\,\textrm{T}$, where $\textsl{g}\approx 2$ is the Land\'{e} \emph{g}-factor, and $\mu_B$ is the Bohr magneton. 
Energy splittings comparable in magnitude have been reported in measurements on similar junctions, but only after applying a magnetic field~\citep{Li_Observation_2013}.

We next apply an in-plane magnetic field ($B$) to the sample. As shown in Fig.~\ref{FirstMag}b, the separation between the  peaks in the $dI/dV$ increases showing a saturation above 30~mT.
The effective spin-splitting increases up to $\sim 190 $~$\mu$eV (which would correspond a magnetic  field of $\sim 1.6$~T), and is preserved even without the presence of an external magnetic field (see the red plot of Fig.~\ref{FirstMag}c).  The superconducting pairing potential $\Delta$ is almost unaffected. 
The enhancement of the spin-splitting can be associated to the increased magnetization of the EuS layer. 

Not only the position, but also the shape of the conductance peaks is different before and after the first magnetization  of the EuS film.  
In the demagnetized phase, the amplitude of inner peaks (at $\mid V \mid < 0.5 $~mV) is larger with respect to the outer ones.  To the best of our knowledge this behavior has  never been reported so far, and  cannot be described by the over-simplified Tedrow-Meservey model that assumes an homogeneous exchange field induced in the superconductor\citep{tedrow_spin-polarized_1986}. 
Below we demonstrate  that this behavior can only be explained by taking into account the multi domain structure of the polycrystalline EuS layer which leads to an inhomogeneous exchange field. 

\begin{figure}
\centerline{\includegraphics[width=1\columnwidth,clip=]{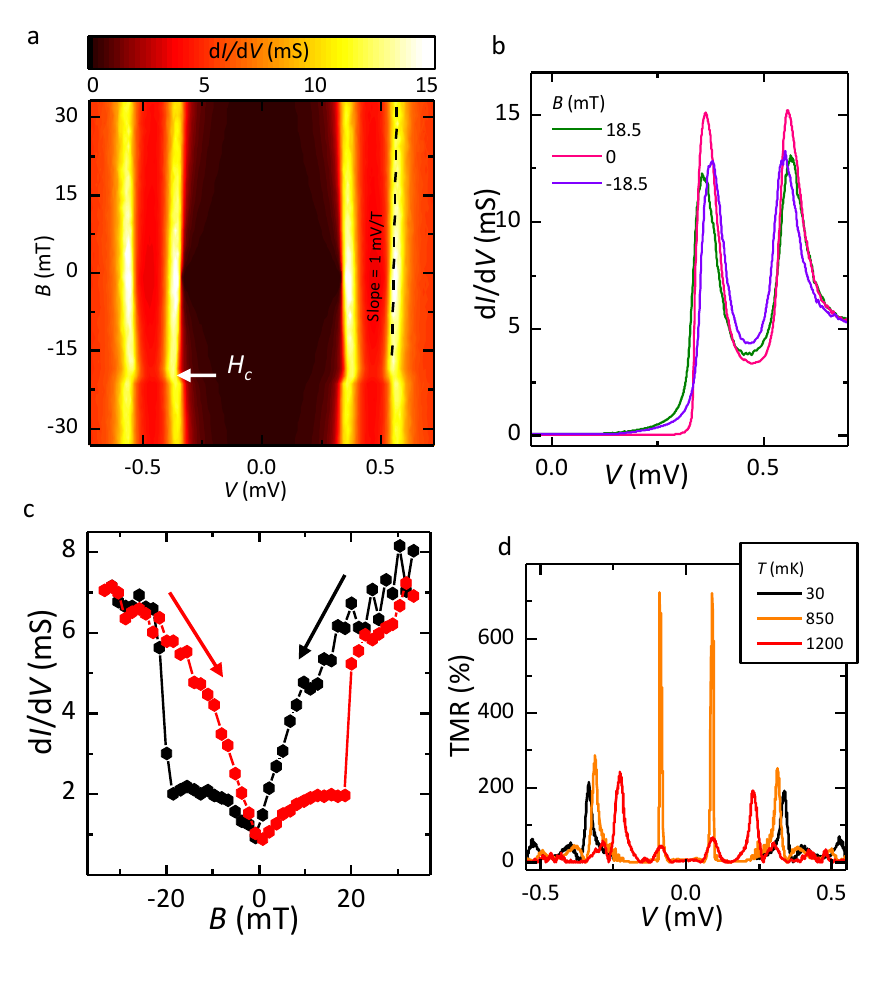}}
\caption{\label{TMR}
\textbf{Hysteretic cicle of the tunnel junction}
(\textbf{a}) Full evolution of the $dI/dV(V)$ of the tunnel junction traced from 30~mT to -30~mT. The dashed line is a guide to the eye following the peak maximum. 
(\textbf{b}) $dI/dV(V)$ for selected values of $B$.
(\textbf{c}) Forward trace ($B: 30 \rightarrow -30$~mT, black dots) and backward trace ($B: -30 \rightarrow 30$~mT, red dots) of the tunneling conductance extrapolated at $V= 335~\mu$V. The measurements of panels a-c were taken at 30 mK.
(\textbf{d}) Tunneling magnetoresistance (TMR) values evaluated at 30, 850 and 1200~mK.
}
\end{figure}

Another striking observation is the sharpness of the tunneling conductance at the gap edge (black curve in Fig.~\ref{FirstMag}c)in the demagnetized phase , in contrast to a smoother transition after the first magnetization (red curve). 
These two different behaviors can be explained by means of the stray field generated by the domain structure of the EuS. In the demagnetized phase the EuS consists of domains with independent magnetization pointing in random directions (see sketch in  Fig. 4a).  Seen from the Al-layer the contributions to the field from different domain walls, being randomly oriented, compensate each other and results in a small stray field.  In contrast in the magnetized  phase, although the number of DWs can be smaller, more domains are aligned and hence their contribution sum up enhancing the stray field. This field acts as pair breaking mechanisms for the superconductor and broaden the BCS peaks. An external magnetic field has the same effect as can be seen in Figs. \ref{TMR}a,b. One clearly sees a larger broadening when a finite field is applied.

\begin{figure}
\centerline{\includegraphics[width=1\columnwidth,clip=]{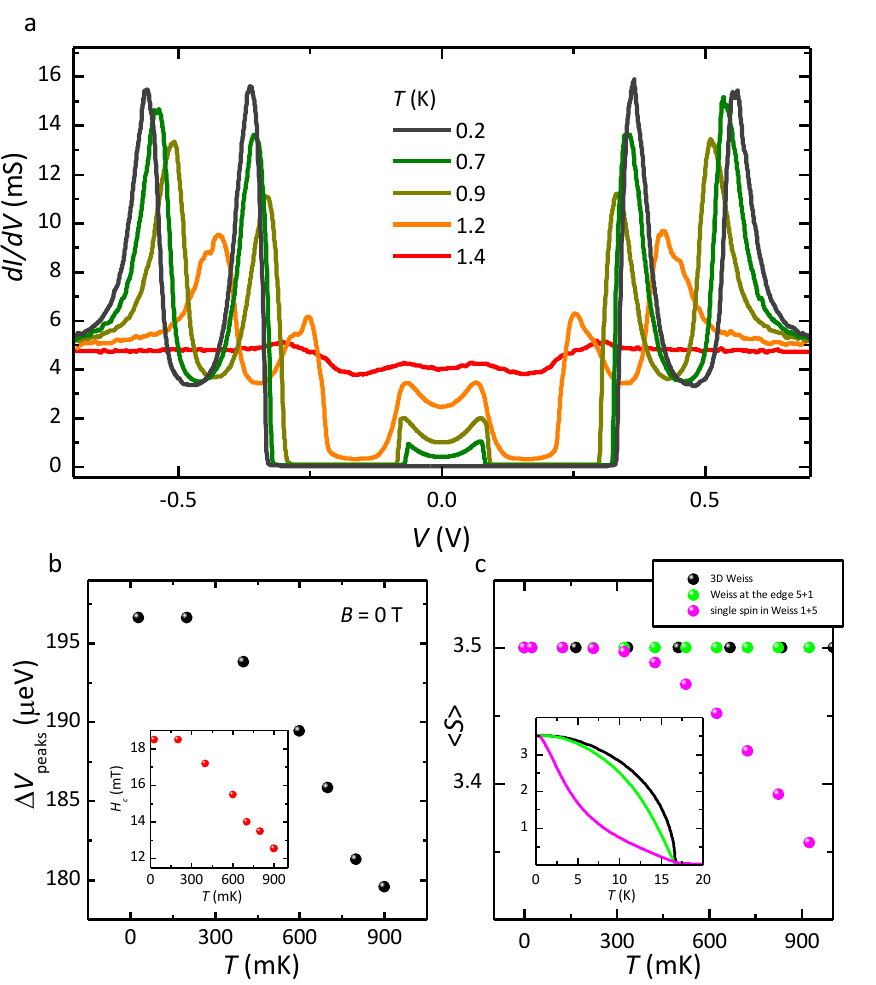}}
\caption{\label{Tdep}
\textbf{Temperature evolution of the the tunnel junction behavior}
(\textbf{a}) Differential conductance $dI/dV(V)$ of the tunnel junction measured at different bath temperatures in zero magnetic field and after the magnetization of the EuS layer.  
(\textbf{b}) Temperature evolution of the exchange splitting extracted form the $dI/dV(V)$ characteristics measured at zero field. Inset: Temperature evolution of the coercive field ($H_c$) extracted form the switching of the tunneling conductance (as shown in Fig.~\ref{TMR}a)
(\textbf{c}) Theoretical temperature dependence of the EuS averaged spin $<S>$ calculated using different approaches. Inset: Evolution of $<S>$ in the full temperature range.
}
\end{figure}

After the first magnetization of the EuS film, the magnetic field dependence of the tunneling  conductance  follows the typical ferromagnetic hysteretic behavior. 
Figure~\ref{TMR}a  shows the typical evolution of the $dI/dV(V)$ extracted from the junction $I-V$  at finite in-plane magnetic fields $B$. The curves show  a clear spin-splitting that increases when the field is applied. This splitting is as large as  $\sim$1mV/T (see  dashed line in Fig.~\ref{TMR}a) and cannot be attributed only to  the Zeeman splitting caused by the external field\cite{xiong_spin-resolved_2011}. The reason  for the large splitting observed  is that  the field tends to enlarge the size of the magnetic domains  and hence  the averaged exchange field seen by the electrons over the Cooper pair size, as explained by our model below.

By reversing the field direction ($B<0$), i.e., anti-parallel to the EuS magnetization, the number and size of the domains with parallel magnetization is reduced. This leads  to  a decrease of the  
the spin-splitting  down to the coercive  $H_{c}\sim -18.5$~mT (see Fig. \ref{TMR}b). 
The discontinuity in the conductance peaks observed at this value of the field is a manifestation of the magnetization switching of the EuS. 
Further increase of the applied field in the negative direction restores  the maximum exchange splitting. By retracing back $B$ a similar hysteretic behaviour is observed  with a coercive field of opposite sign ($H_{c}\sim 18.5$~mT).

The junction hysteretic behaviour joined to the strong quenching of the differential conductance at sub-gap voltages ($|e V| < 2 \Delta - h_{ex}$) suggest the possibility to operate this structure as a magnetic switching device.
Notably, such a device is based just on a single ferromagnetic layer and could, in principle, be exploited as a \emph{permanent} memory element. The performance of the junction as a non-volatile memory can be quantified by its tunneling magnetoresistance (TMR) evaluated from the hysteretic spectra of the $dI/dV(B)$ curves shown in Fig.~\ref{TMR}c, and defined as
$$TMR = Max(G_{fw}/G_{bk},G_{bk}/G_{fw})-1$$ where $G_{fw}$ ($G_{bk}$) is the forward (backward) differential conductance.
As shown in Fig.~\ref{TMR}d, the TMR at 30~mK can exceed 200$\%$ by tuning the bias voltage in the subgap energy regime, $V\simeq 335 \mu$V, corresponding to the active  voltage range for which the junction is switching between the insulating state (i.e., in the sub-gap conductance) and the conducting state according to the magnetic configuration of the EuS. 
Furthermore, the figure shows that such a  high TMR value is preserved as well by increasing the bath temperature up to $T\lesssim T_c$ as, in this temperature window, thermal broadening  is negligible compared to the energy scale of the exchange splitting. In addition, at higher bath temperatures (i.e., for $T > 0.5$~K) the TMR shows an interesting additional feature in the sub-gap region (around $V\simeq 80 ~ \mu$V) that is even more sensitive to the magnetic switching (TMR$> 700\%$).
This sub-gap feature stems from the presence of the superconducting matching peaks which are activated by the temperature and exchange field in these junctions.
These can be appreciated in Fig.~\ref{Tdep}a showing the differential conductance $dI/dV(V)$ measured at different temperatures. As predicted by the quasiparticle spin-polarized tunneling theory, temperature enhances the subgap matching peak expected at $|e V| \simeq \Delta_1 - \Delta_2 + h_{ex} \simeq 80~\mu$V. The position of these additional maxima provides an alternative estimation of the  energy splitting that is consistent with the BCS peaks splitting observed at higher voltages.

From the tunneling conductances measured at different temperatures we extracted the temperature evolution of both, the exchange energy and the coercive field in a region of temperatures never explored so far for EuS. These results are presented in Fig.~\ref{Tdep}b. Both the exchange energy and the coercive field increase by lowering the temperature, suggesting that the ferromagnetic ordering of the EuS and, in turn  the resulting magnetic proximity effect,  are affected even in a temperature range much lower than the Curie temperature of the EuS. As discussed below,  this anomalous behaviour can be explained to originate from the magnetic properties of weakly coupled  Eu atoms at the interface with the Al layer. 

\section{Theoretical description of the magnetic proximity effect}
\begin{figure*}[t!]
\centerline{\includegraphics[width=1.0\textwidth,clip=]{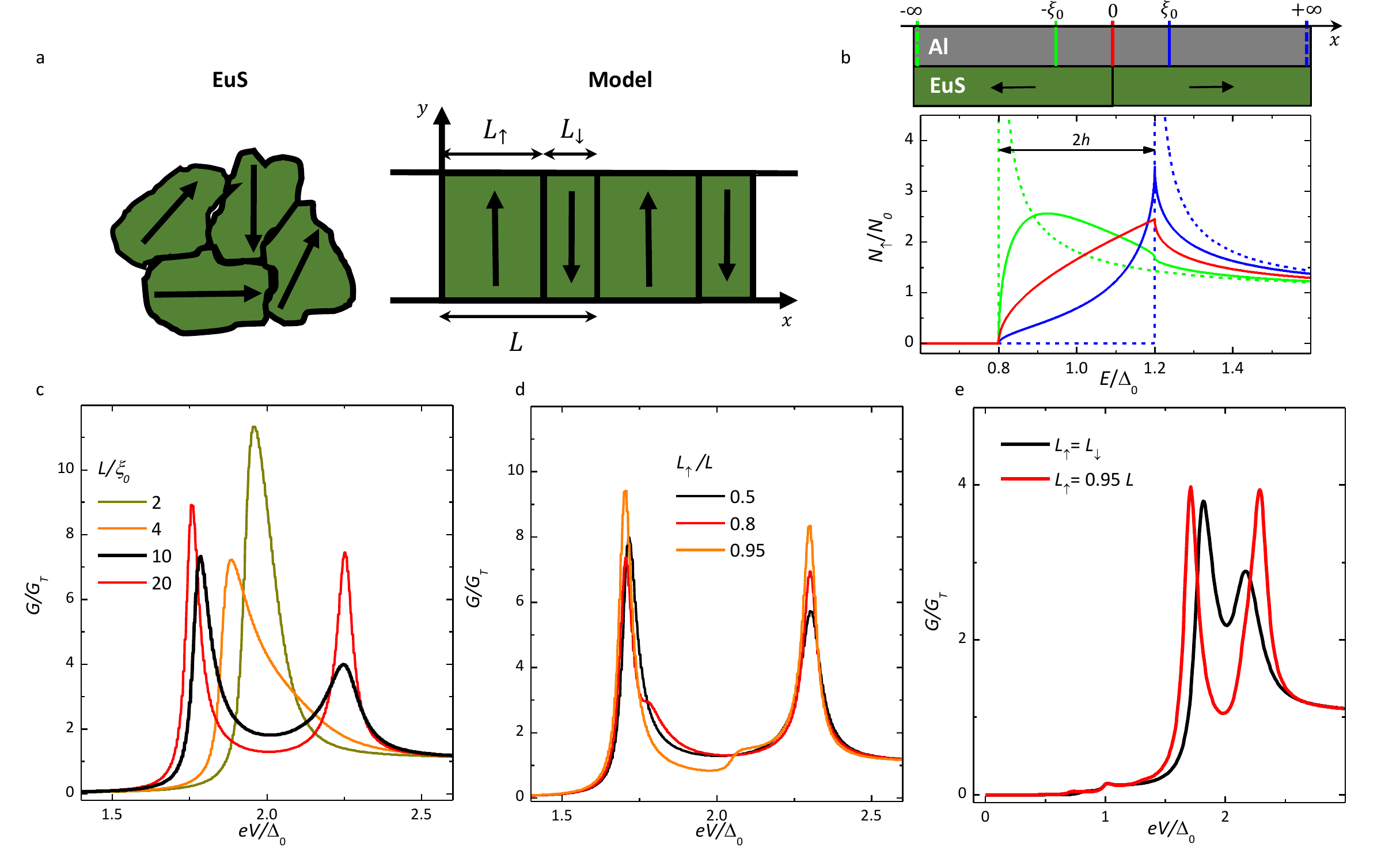}}
\caption{
\label{FigTheo}
\textbf{Theoretical Model of the tunneling conductance}
(\textbf{a}) left panel: sketch of the polycrystalline structure of the EuS, right panel: effective model of alternating up down domains used in the simulation. We consider two lengths  $L_\uparrow$ and $L_\downarrow$ and define $L= L_\uparrow + L_\downarrow$.  The demagnetized phase is described by $L_\uparrow=L_\downarrow$, whereas after  magnetization $L_\uparrow\gg L_\downarrow$.
(\textbf{b}) Local density of states of the Aluminum on top of a two semi-infinite domain structure. The dashed lines show the spin-splitting BCS DoS deep inside the domains. at $x\to-\infty$ (blue) and $x\to+\infty$ (green). By approaching the domain wall the DoS changes.  Line traces of the plot taken at $x=-\xi_0$ (blue), $x=0$ (red), and $x=\xi_0$ (green).
(\textbf{c}) Averaged tunneling conductance calculated for a infinite stripe of balanced up/down domains $L_\uparrow= L_\downarrow$ then describing the demagnetized EuS calculated for different domain size ($L$), $\Gamma = 0.01$ and $T= 0.01 T_C$.
(\textbf{d}) Evolution of the tunneling conductance calculated at different magnetizations $L_\uparrow$ and $L=10 \xi_0$, $\Gamma = 0.01$ and $T= 0.01 T_C$.
(\textbf{e}) Tunneling conductance calculated for a demagnetized phase, made of six different domains, (black plot)  and for a fully magnetized phase (red plot) including magnetic scattering. 
+}
\end{figure*}
In all previous works on EuS/Al,  the spin-splitting  is modeled  by assuming an homogeneous exchange energy $h_{ex}$.
In such a situation, the DoS of the Al-layer  can be approximated  by  the sum of the  DoS for spin up and spin down:
\begin{equation}
N_s^{hom.}=\frac{1}{2}\sum_{\sigma=\pm1}N_{BCS}(E+\sigma h_{ex})\; ,  \label{eq:hom_h}
\end{equation}
where $N_{BCS}(E)=\left|\Real[(E+i\Gamma)/\sqrt{(E+i\Gamma)^{2}-\Delta^2}]\right|$ is the usual BCS DoS and $\Gamma > 0$ is the Dynes parameter describing inelastic scattering. The exchange energy  $\pm h_{ex}$ describes the splitting  observed in the tunneling conductance of the FI/S-I-S tunnel junctions, and the shape of the dI/dV(V) derived from Eq. (\ref{eq:hom_h}) is very similar to the one observed in the magnetized case (see for example red  curve in Fig.~\ref{FirstMag}c). 

In the demagnetized case, although the total magnetization
is negligibly small, a clear splitting is observed in the experiment ( black curve in  Fig.~\ref{FirstMag}c).  However, 
the shape of the dI/dV(V)  curve is very different from the one observed after the first magnetization of the EuS, and hence cannot be described by the homogeneous  DoS given by   Eq. (\ref{eq:hom_h}). In other words, if one would assume 
 that the enhancement of the splitting  after  magnetizing the EuS layer is due to the increase of an homogeneous exchange $h_{ex}$ in Eq. (\ref{eq:hom_h}), one  would not be able to explain the reversed relative height of of the inner and outer  peaks  in Fig.~\ref{FirstMag}c. 
 
The  main assumption for our theoretical model is that the EuS consists of magnetic domains, typical for ferromagnetic materials. Indeed, it is known  that EuS films are polycrystalline and consist of an ensemble of crystallites with intrinsic magnetization \cite{tischer_ferromagnetic_1973}. 
In the absence of an applied field and before the first magnetization,  each crystallite can be 
regarded as a single domain magnet with their ensemble having random magnetization orientation relative to each other (see Fig.\ref{FigTheo}a). The typical size of such domains can be of  the order of several hundreds of nanometers, and depends
on different factors as for example, temperature, growing conditions,etc.
Because of a weak anisotropy, when a  magnetic field is applied, the magnetic moments of the crystallites tends to orient themselves parallel to the applied field, forming large domains leading to homogeneous magnetization.

The spin-splitting  observed in the differential conductance of  FI/S-I-S junctions are  attributed to the 
magnetic proximity effect, {\it i.e} to the exchange interaction between the spin moment of the Eu ions ${\bf S}$ and the spin density of conducting electrons ${\bf s(r)}$. To describe this interaction we assume a simple exchange Hamiltonian:
\begin{equation}
H_{ex}=-J\sum_j\bm{S}_j\cdot\bm{s}\left(\bm{r}_j\right),
\label{TeqHeisenbergHJ}
\end{equation}
where $J$ is the interfacial coupling constant. By averaging Eq.~(\ref{TeqHeisenbergHJ}) in the ferromagnetic state of the EuS, we obtain the local exchange coupling
\begin{equation}
\hat{h}_{\textit{ex}}(x,y,z)=-\frac{1}{2}
J n_{\textrm{2D}}\hat{\bm{\sigma}}\cdot\bm{S}(x,y)\delta(z-z_M)\; .
\label{Teqhexmicr0}
\end{equation}
Here $n_{\textrm{2D}}$ is the two-dimensional concentration of Eu ions accessible to the Al electrons at the interface, $\bm{S}(x,y)$ is the average value of the interfacial local moment, $\hat{\bm{\sigma}}$ are the Pauli spin matricies and $z_M$ is the position of the EuS/Al interface. 
By assuming such effective exchange interaction the spectrum of the superconductor adjacent to the EuS can be determined from the quasiclassical Green's function, $\check{g}$, that in the diffusive limit  satisfies the Usadel equation~\cite{Usadel_Generalized_1970,bergeret_odd_2005}:
\begin{equation}
\hbar D {\bf \nabla}
\check{g} {\bf \nabla} \check{g}
+ i E\left[\tau_3,\check{g}\right]
- i \left[\tau_3\hat{h},\check{g}\right]
+ \Delta\left[\tau_2,\check{g}\right]=\left[\check\Sigma,\check g\right] ,
\label{TeqUsadel0}
\end{equation}
under the constraint $\check{g}^2=\check{\mathbb{1}}$ and with suitable boundary conditions (details  of the notations  and the boundary problem  are given in the Methods section).  The exchange energy $\hat{h}$ entering this equation consists of the Zeeman term and an interfacial exchange term,
$\hat{h}(\bm{r})=
\frac{1}{2}\textsl{g}\mu_B \hat{\bm{\sigma}}\cdot\bm{B}
+\hat{h}_{\textit{ex}}(\bm{r})$. 
The term in the r.h.s of (\ref{TeqUsadel0}) , describes  possible sources for inelastic scattering or pair-braking mechanisms describes by the self-energy $\check \Sigma$.  In the most simple case one describes inelastic scattering by the energy independent Dynes parameter such that $\check\Sigma=\Gamma \tau_3$.   

Equation (\ref{TeqUsadel0}) determines the length scales over which the spectral properties of the Aluminum  are modified. This scale  is of the order $\sim \xi_0 = \sqrt{\hbar D/ \Delta} $.
If we assume that the thickness of the Al layers is small compared to this length we can integrate Eq. (\ref{TeqUsadel0}) over the thickness ($z$-direction), and reduce the 3D to a 2D problem (details in Methods).  Specifically, given a magnetic configuration, ${\rm S}(x,y)$ of the EuS at the interface $z=z_M$ one has to solve Eq. (\ref{TeqUsadel0}) to  determine the local DoSs of the Al film from the equation: 
\begin{equation}
N_\sigma(E,x,y)=\frac{N_0}{2}\Real\left[\Tr\left\{g_\sigma(E+i0,x,y)\tau_3\right\}\right],
\end{equation}
where $\Tr\{\dots\}$ stands for the trace in the Nambu space. 
It is easy to check that in the homogeneous case the solution of the Usadel equation gives  the simple spin-split BCS DoSs of Eq. (\ref{eq:hom_h}). In order to describe the polycrystalline  phase of the EuS  with random magnetization we model it by assuming a stripe of domains with alternating up and down magnetization (see Fig.  \ref{FigTheo}a).  The relative size between up and down domains will depend on the magnetization state of the EuS. This reduces further the problem to 1D. As we shall see, even with this simplification we are able to catch most of the experimentally observed conductance features.

It is instructive to focus first  on the problem of two semi-infinite domains with opposite magnetization and a single domain wall at $x=0$. This case in fact can be solved analytically.  In Fig.  \ref{FigTheo}b we show the local  DoS  (for one spin species) calculated from the   Usadel  equation at different points. 
As expected,  deep in the bulk of the domains, ($\left|x\right|\gg\xi_0$) the density of states is a BCS peak shifted on each side of the domain wall , $N_\uparrow(E,\pm\infty)=N_{\textrm{BCS}}(E\pm h)$(dashed lines).
The BCS singularity is recovered only asymptotically with $x\to\pm\infty$. On a length scale $\xi_0$ around the domain wall  there is a crossover from the two shifted BCS curves to a "shark-fin" shape at the domain wall, $x=0$. 
It is important to emphasize that the inner "peak" at $E=\Delta-h$ looks as being shifted to larger energies when moving towards the domain wall, whereas the  features at $E=\Delta+h$  remain at the same energy.  
These features, that will be used below to understand the dI/dV curves,  could be verified by measuring the local density of states  with for example a scanning tunneling microscope{\cite{moussy_local_2001,le_sueur_phase_2008}.
 Here we perform a planar  tunneling spectroscopy, with a large contact area between the FI/S bilayer and the superconducting electrode (see Fig.~\ref{FirstMag}a). 
 This means, in particular, that by  measuring the tunneling differential conductance of the junctions we obtain information about the DoS averaged over the area of the tunnel barrier and the two spin species ($ \bar{N}_{}(E) =\sum_{\sigma} < N_{\sigma}(E,\sigma,x)>_x $ ):
\begin{equation}
\frac{dI}{dV}(V)=\frac{G_T}{e}\frac{d}{dV}\int dEN_{BCS}(E+eV)\bar{N}_{}(E)\left[f(E)-f(E+eV)\right]\;,\label{eq:didv}
\end{equation}
where $G_T$ is the normal-state conductance of the tunneling barrier, and $f(E)$ is the Fermi function. 
To extend the model to a realistic multi-domain structure we solved numerically the Usadel equation, and calculated the average DoS for an infinite stripe (see Fig.~\ref{FigTheo}a) made of two domains of length $L_\uparrow$ and $L_\downarrow$ repeated with $L=L_\uparrow+L_\downarrow$ periodicity. The ratio  $L_\uparrow/L_\downarrow$  determines 
 the total magnetization of the EuS. In the
demagnetized phase $L_\uparrow/L_\downarrow=1$, whereas after the magnetization we assume $L_\uparrow/L_\downarrow\gg1$. The other important parameter of the theory is the ratio $L/\xi_0$ that, as we see below,   determines crucially the shape of the dI/dV(V) curves obtained by the  tunneling spectroscopy.

For  the demagnetized phase of EuS we assume that  $L_\uparrow=L_\downarrow$ and explore 
 the role of the domain size on the tunneling conductance. This is shown in Fig. \ref{FigTheo}c, where  $dI/dV(V)$ curves are shown for different values of   $L/\xi_0$. Despite the fact that the total magnetization of  the EuS  is zero, a clear splitting is visible for large domains $L>4 ~\xi_0$ and  reaches the asymptotic value $2 h_{ex}$ above $20~\xi_0$.  These results suggest a typical domain size of $\sim 10 ~\xi_0$ in the EuS films. Moreover, our model also described correctly  the relative heights  of the peaks in the demagnetized phase. 
 
After applying the magnetic field  the ratio $L_\uparrow/L_\downarrow$ is increased.  By fixing   the period of the structure $L=L_\uparrow+L_\downarrow=10\xi_0$ we show in Fig. \ref{FigTheo}, the $dI/dV(V)$ for different values of $L_\uparrow/L_\downarrow$. Our results clearly show that  the separation between the spin-split peaks increases by increasing the ratio $L_\uparrow/L_\downarrow$ which 
modifies also the relative heights of the peaks that tends
to the case of homogeneous field result at $L_\uparrow/L_\downarrow \rightarrow \infty$ . 
Despite the fact that the  model assumes a unique domain size for each spin species most features of  Fig. \ref{FirstMag} are caught within this model.

There are however two main discrepancies between the theoretical results and the measurements: On the one hand the peaks observed experimentally show a much larger broadening than seen in the calculated ones. This discrepancy is easy to understand recalling that in a real  situation magnetic-disorder, spin-orbit coupling  and the effect of  stray fields will broaden all the features\cite{meservey_spin-polarized_1994}. These effects are energy dependent (see r.h.s of Eq. (\ref{TeqUsadel0})] and for simplicity have not been included in the simulation. Instead, we modeled the   inelastic scattering  by the energy independent  Dynes parameter  $\Gamma=0.01\Delta_0$.
On the other hand there is a more important discrepancy 
if  one compares  the results of  our model, Fig.  \ref{FigTheo}d, with the  measurements before and after the first magnetization, Fig. \ref{FirstMag}c.
In the latter we clearly see that by magnetizing the EuS layer the splitting peaks move symmetrically with respect to the voltage $eV=2\Delta_0$. In contrast, our simulations, Fig. \ref{FirstMag}c, shows that only the inner peak is shifted by changing  the value $L_\uparrow/L_\downarrow$.  The voltage at which the  outer peak  appears  does not change though, in accordance with the result for the DoS using the two infinite domains model, Fig.  \ref{FigTheo}b .

The latter discrepancy is a consequence of the assumption we made   that  the sizes of all up and down domains is unique.  In reality, the size of the domains  follows certain distribution with  an average domain size  given, according to our previous results to  $\sim10\xi_0$.  In order to describe this situation within our model, 
we assume for example that inside the up(down) domain there is a smaller down(up) domain.
 Fig. \ref{FigTheo} e shows the resulting  tunneling conductance obtained  by assuming small domains with a size 10\% of the host domains.   
  The effect of the small domains embedded in  the larger ones  is to reduce the magnitude of the effective spin-splitting that leads to the symmetric  shift of the  peaks.  In order to  broaden  the  peaks we have  used  a larger  Dynes parameter , $\Gamma=0.03\Delta_0$. Being  independent of the energy  its inclusion  leads to  sub-gap features  which can be neglected.

\section{ Temperature dependence of the exchange coupling}

The last  striking  feature to be explained  is the strong temperature dependence of the exchange coupling observed in  Fig.  \ref{Tdep}b. The surprising issue is the small temperature window over which the exchange field changes even at such low temperatures. This energy scale is clearly not related to the Curie temperature of the EuS layer, which is more than one order of magnitude larger.
A similar feature was also reported in Ref. \cite{xiong_spin-resolved_2011} when the sample was immersed in a magnetic field of 50 mT, and was attributed to a "thermally activated" spin-relaxation mechanism, although the authors did not elaborate this hypothesis.
We provide here an alternative and more plausible explanation.
  
According to our description of the magnetic proximity effect,  the exchange field is proportional to the average (localized) spin, Eq. \ref{Teqhexmicr0}.   In order to estimate  this average we have calculate the magnetization for a cubic lattice with S=7/2 in the nodes and with Heisenberg exchange interaction between nearest neighbors. 
For the exchange coupling of $J=0.0688$ meV, we can  recover $T_C=16.7$ K, using the self-consistency equations of the RPA theory. This calculation (see Fig.~\ref{Tdep}c)  demonstrates that the change of magnetization in going from 30 mK to 900 mK is negligibly small ($< 0.4 \%$) in the bulk of the EuS film.
But at the surface, the effect might be somewhat larger, because the spins have 5 nearest neighbors and not 6 as in the bulk. 
To verify how large this change is, 
 we compute within the Weiss mean-field theory  the surface average spin.   
 The change becomes $10^{-6}$ in the usual Weiss theory, and $10^{-5}$ in the relaxed at the surface Weiss theory. 
Thus, in both cases  the average spin do not have any special characteristic scale other than the usual Curie temperature, and therefore one cannot explain the change on $h_{ex}$ observed in Fig.  \ref{Tdep}b  from the ferromagnetic phase of plain EuS.   
   
An alternative explanation is  that the  observed 10$\%$ reduction of the effective splitting  in going from 25 to 900 mK could be attributed to the increase of $\xi_0$, and hence to the  reduction of the averaged exchange field.   This could be correct provided the $B=0$ data in  Fig.  \ref{Tdep}b was taken "before" magnetization.  But this is not the case. Moreover, in Ref. \cite{xiong_spin-resolved_2011}  the same behaviour was observed in the presence of a large magnetic field 

In order to understand this issue we propose the following scenario: Most likely, the EuS surface has a portion of spins which do not have the 5+1 coordination (here 1 stands for the Al atom). 
There should be spins which stand out of the lattice and are coupled to the rest of the EuS by just one single bond with the same exchange $J=0.0688$ meV as the bulk Eu spin.  These loose surface spins  correspond to a  1+5 coordination (now 5 stands for Al atoms and 1 for Eu).  For such  spins, we get  the cyan curve in the plot of Fig.~\ref{Tdep}c. The new characteristic temperature scale of the down bending of the curve is basically given by $J$, which could in principle be even smaller than that for the lattice. This explains the change of the average spin, and thereby of the effective exchange field over such a small temperature window.

%

\section{Conclusions}

In summary, by combining tunneling conductance measurements and a microscopic model based on the quasiclassical Green's functions we provided an exhaustive description of the magnetic proximity effect in ferromagnetic insulator/superconductor EuS/Al bilayers.  
We identified two different magnetic scenarios which change whether the system was first magnetized or not. 
By comparing our calculations of the density of states and tunneling conductance with the measurements we conclude that the EuS film consists of crystallite with sizes (and domains)  larger than the superconducting coherence length.  In the demagnetized phase of  the EuS layer, each of these crystallites has an independent magnetic moment randomly oriented. We modeled such crystallites as magnetic domains that caused an  exchange field     parallel  to the  local average magnetization. 
 Because of the large mean size of the domains in comparison to the coherence length of the superconductor, even before applying any magnetic field the spectrum of the Al layer shows a well defined  spin-splitting. 
 
By applying a magnetic field the magnetization of the crystallites start to form larger  magnetic meta-domains with an homogeneous magnetization parallel to the applied field.  This manifests as an enhancement of spin-splitting  of the density of states of the superconductor and a modification of $dI/dV(V)$ curves towards the ones assumed in previous works for an  ideal homogenoeus magnetization. Moreover, the observed spin-splitting, evolving in a temperature range much smaller than the Curie temperature of the bulk EuS, reveals the
presence of weakly bound spins at the interface of the EuS/Al. 

Because of  the large spin-splitting observed even   in the absence of any applied magnetic field, 
the EuS/Al material combination is  an excellent platform for the development of devices requiring the coexistence of superconducting correlations and spin-splitting exchange fields, as for example in the field of  Majorana- based  quantum computation .
 
\section{Acknowledgements}

Partial financial support from the European Union's Seventh Framework Programme (FP7/2007-2013)/ERC Grant 615187-COMANCHE is acknowledged. The work of E.S. is funded by the Marie Curie Individual Felloship MSCA-IFEF-ST No. 660532-SuperMag. The work of G.D.S. is funded by Tuscany Region under the FARFAS 2014 project SCIADRO. The work of F. S. B. and V. G. was supported by Spanish Ministerio de Economía y Competitividad (MINECO) through Project No. FIS2014-55987-P. J.S.M. acknowledges the support from NSF Grants No. DMR-1207469, ONR Grant No. N00014-16-1-2657 and John Templeton Foundation grant.

\section{Methods}

\subsection{Sample fabbrication}
The structure of the magnetic tunnel junctions  investigated is EuS(4)/Al(4)/Al$_2$O$_3$/Al(8) (the thickness is in nanometers), where materials are listed in the order in which they were deposited. The junctions were fabricated in a vacuum chamber with a base pressure $2\times 10^{-8}$Torr using in situ shadow masks. To facilitate the growth of smooth films, a thin Al$_2$O$_3$ (1 nm) seed layer was deposited onto chemically and further in situ oxygen-plasma cleaned glass substrates.
The substrate was cooled to liquid-nitrogen temperature for the growth of EuS and Al layers. After warming to room temperature, a thin Al$_2$O$_3$ barrier was formed by plasma oxidization of the 4 nm Al film surface. The top Al film was deposited at room temperature over this, and then the junctions were capped with 6 nm of Al$_2$O$_3$ for protection. 


\subsection{Brief description of the theory}

We describe the electronic properties of the superconducting film with the help of the quasiclassical Green function $\check{g}(\bm{r})$, obtained as a solution of the Usadel, Eq. ({TeqUsadel0}), equation~\cite{Usadel_Generalized_1970,bergeret_odd_2005}.
We described  inelastic scattering  by  energy independent  Dynes term  in the r.h.s 
\begin{equation}
\frac{D}{2}\sum_{\alpha}\partial_\alpha
\left[\check{g},\partial_\alpha\check{g}\right]
+ i E\left[\tau_3,\check{g}\right]
- i \left[\tau_3\hat{h},\check{g}\right]
+ \Delta\left[\tau_2,\check{g}\right]=[\Gamma\tau_3,\check g]\; .
\label{eqUsadel0}
\end{equation}
Here, $D$ is the diffusion constant, $E$ is the quasiparticle excitation energy, $\Delta$ is the superconducting gap, and the sum runs over the spacial directions ($\alpha=x,y,z$). We use two sets of Pauli matrices, $\bm{\tau}$ and $\hat{\bm{\sigma}}$, to represent quantities in the Nambu and spin spaces, respectively. A check accent ($\check{g}$) denotes a $4\times 4$ matrix in the direct product of the
spin and Nambu spaces, whereas a hat accent ($\hat{h}$) denotes a $2\times 2$ matrix in the spin space only. 
The exchange field  $\hat{h}(\bm{r})$  consists of the Zeeman term and an interfacial exchange term 
\begin{equation}
\hat{h}(\bm{r})=
\frac{1}{2}\textsl{g}\mu_B \hat{\bm{\sigma}}\cdot\bm{B}
+\hat{h}_{\textit{ex}}(\bm{r}),
\end{equation} 
where $\textsl{g}\approx 2$ is the Al g-factor, $\mu_B$ is the Bohr magneton, $\bm{B}$ is the magnetic field,
and $\hat{h}_{\textit{ex}}(\bm{r})$ is the exchange field coming from the magnetic interface that in general  is inhomogenoues in  space.

The boundary conditions for $\check{g}$ inside the Al film at its upper ($z=0$) and lower ($z=-d$) surfaces 
are obtained by infinitesimal integration across each interface, assuming that $\check{g}$ vanishes identically 
both in the $\textrm{Al}_2\textrm{O}_3$ barrier ($z>0$) and in the ferromagnetic insulator EuS ($z<-d$),
\begin{equation}
-\frac{D}{2}\left[\check{g},\partial_z\check{g}\right]=
\left\{\begin{array}{ll}
0,&\quad\quad z=0,\\
-i \left[\tau_3\hat{v},\check{g}\right],&\quad\quad z=-d,
\end{array}\right.
\label{eqbcs12}
\end{equation}
where $\hat{v}(x,y)=-\frac{1}{2}J n_{\textrm{2D}}\hat{\bm{\sigma}}\cdot\bm{S}(x,y)$.
Thus, the polarization of the superconductor is intimately connected with the magnetic structure of the EuS film through the quantity $\hat{v}(x,y)$ in Eq.~(\ref{eqbcs12}).

Despite the fact that the exchange field $\hat{h}_{\textit{ex}}(\bm{r})$ is strongly localized at the lower surface of the Al film, the tunneling density of states probed on the upper surface is modified by $\hat{h}_{\textit{ex}}(\bm{r})$ equally strongly as on the lower surface, provided $d$ is small compared to the superconducting correlation length. 
Specifically, if  $\hat{v}\ll D/d$ we find that the Green function on the upper surface ($\check{g}_0$) satisfies a 2D version of the Usadel equation, which differs from Eq.~(\ref{eqUsadel0}) only by a reduced dimensionality ($\alpha=x,y$) and an effective exchange field 
$\hat{h}\to \hat{h}_{\textrm{eff}}(x,y)$
%
The magnetic structure of EuS can, therefore, be probed through a relatively thick Al layer ($d\lesssim\xi_0$), by studying, \emph{e.g.}, the superconducting density of states at excitation energies $E\lesssim\Delta$. 
Strictly speaking in the vicinity of sharp domain walls o a step-like change of $\hat{v}(x,y)$ is imaged on the upper Al surface as a gradual transition over a length scale $d$.

\bibliographystyle{apsrev4-1}
\bibliography{EliaLibrary2}

\end{document}